\documentclass[preprint,aps,amsmath,amssymb]{revtex4}
\usepackage{graphicx}
\usepackage{bm}

\begin{document}


\title{Formation of mid-gap states and ferromagnetism in semiconducting CaB$_{6}$}
\author{Jong-Soo Rhyee$^{1}$, B. H. Oh$^{1}$, B. K. Cho$^{1}$*, M. H. Jung$^{2}$, H. C.
Kim$^{2}$, Y. K. Yoon$^{3}$, Jae Hoon Kim$^{3}$, and T.
Ekino$^{4}$} \affiliation{$^{1}$Center for Frontier Materials,
Department of Materials Science and Engineering, Kwangju Institute
of Science
and Technology, Gwangju 500-712 Korea\\
$^{2}$Materials Science Lab. Korea Basic Science Institute, Taedok
Science Town, Taejon 305-333 Korea\\
$^{3}$Institute of Physics and Applied Physics, Department of
Physics, Yonsei University, Seoul 120-749 Korea\\
$^{4}$Faculty of Integrated Arts and Sciences, Hiroshima
University, Higashi-Hiroshima 739-8521, Japan\\
$^*$Correspondence should be addressed to the following E-mail:
chobk@kjist.ac.kr (B.K.C.)}
\date{\today}

\begin{abstract}
We present a consistent overall picture of the electronic
structure and ferromagnetic interaction in CaB$_6$, based on our
joint transport, optical, and tunneling measurements on
high-quality {\em defect-controlled} single crystals. Pure CaB$_6$
single crystals, synthesized with 99.9999 \%-pure boron, exhibited
fully {\em semiconducting} characteristics, such as monotonic
resistance for 2--300 K, a tunneling conductance gap, and an
optical absorption threshold at 1.0 eV. {\em Boron-related
defects} formed in CaB$_6$ single crystals synthesized with 99.9
\%-pure boron induced {\em mid-gap states} 0.18 eV below the
conduction band and extra free charge carriers, with the
transport, optical, and tunneling properties substantially
modified. Remarkably, no ferromagnetic signals were detected from
single crystals made with 99.9999 \%-pure boron, regardless of
stoichiometry, whereas those made with 99.9 \%-boron exhibited
ferromagnetism within a finite range of carrier density. The
possible surmise between the electronic state and magnetization
will be discussed.

\end{abstract}
\maketitle

The hexaboride compounds {\it R}B$_6$ ({\it R} = Ca, Sr, La, Ce,
Sm, Eu, and Gd) have been studied extensively over the last few
decades because of their distinctive physical properties.
Recently, it was found that CaB$_6$ doped with a small amount of
La exhibits a weak ferromagnetism at a high temperature ($T_c$
$\approx$ 600 K) without the constituent elements being partially
filled with $d$ or $f$ orbitals, which are usually required for
ferromagnetism \cite{Young}. Substantial theoretical and
experimental efforts have been devoted to clarify this intriguing
property \cite{Zhitomirsky,Ceperley,Jarlborg,Massida,Tromp}. Based
on the early band structure calculations of CaB$_6$
\cite{Massida}, from which the compound's apparent semimetallic
character could be derived, many theoretical models, such as the
ferromagnetic phase of a dilute electron gas (DEG), the doped
excitonic insulator (DEI), and the conventional itinerant
magnetism, have been suggested to explain the weak ferromagnetism
observed at a high temperature
\cite{Zhitomirsky,Ceperley,Jarlborg}. However, a more detailed
calculation employing the so-called {\it GW} approximation
\cite{Tromp} predicted that CaB$_{6}$ has a sizeable
semiconducting band gap of about 0.8 eV at the \textit{X} point in
the Brillouin zone and suggested that the magnetism in
Ca$_{1-x}$La$_x$B$_6$ occurs just on the metallic side of a Mott
transition in the La-induced impurity band. Experimental
investigations of the predicted band gap with angle-resolved
photoemission spectroscopy (ARPES) supported the results of the
{\it GW} calculation \cite{Denlinger,Souma}. The mapped band
structure and the Fermi surface were in good agreement with the
\textit{GW} scheme, and a semiconducting band gap at the
\textit{X} point, estimated to be 1 eV \cite{Souma} or larger
\cite{Denlinger}, was reported. In addition, a small electron
pocket found at the \textit{X} point was thought to originate from
boron vacancies and to cause the previously reported metallic
conductivity in CaB$_{6}$, SrB$_{6}$, and EuB$_{6}$. A more direct
signature of a semiconducting band gap, such as in optical
absorption and tunneling conductance, would be highly desirable.

Regarding the magnetic properties of La-doped CaB$_6$, there have
been a lot of debates on the origin of the magnetic moment and on
the mechanism of the exotic ferromagnetism. Recently, FeB and
Fe$_{2}$B, which have critical temperatures of about 598 K and
1,015 K, respectively, were suggested to be responsible for the
high-temperature weak ferromagnetism in CaB$_6$
\cite{Matsubayashi}, while the evidence against the argument was
presented by Young \textit{et al.} on an experimental basis
\cite{Young02}. However, it was also suggested that defects,
possibly driven by La doping and randomly distributed in the
lattice, generate free charge carriers that simultaneously create
local magnetic moments \cite{Fisk}. The defects likely involve
sites in the boron sublattice rather than those in the cation
sublattice since excess Ca or La ions are not likely to be placed
into the rigid network of B octahedra.  Indeed, the formation
energy and local magnetic moment of a series of possible point
defects in CaB$_{6}$ were calculated \cite{Monnier}. However, the
exact nature of the hypothetical defects is still not clear and it
should be confirmed experimentally, in particular whether the
defects really induce local magnetic moments and possibly free
charge carriers as well.

We synthesized our CaB$_6$ single crystals using the
high-temperature flux method, described in detail elsewhere
\cite{Ott}. In order to initiate a variation in the relative
concentration of Ca and B, we started the single-crystal growth
process with the initial molar ratios of Ca:B = 1:5, 1:6, and
1:6.05. We denote the B-deficient single crystal as
CaB$_{6-\delta}$, the stoichiometric one as CaB$_{6}$, and the
B-rich one as CaB$_{6+\delta}$, depending on the initial molar
ratio of Ca and B. For the growth of each single-crystal species,
we used two different types of boron, one with 99.9999 \% purity
(6N) (EAGLE PICHER, USA) and the other with 99.9 \% purity (3N)
(Target Materials Inc, USA). In the following, a single crystal
denoted as CaB$_{6-\delta}$(6N) is a nominally B-deficient sample
made with 99.9999 \%-pure boron, and likewise CaB$_{6-\delta}$(3N)
for one made with 99.9 \%-pure boron. Similar notations are used
for the stoichiometric and the nominally B-rich single crystals.
The 6N and 3N boron contain Fe impurities of order of 0.01 ppm/mg
and 1.0 ppm/mg, respectively, and other major non-magnetic ones of
C, and Si was found in 3N boron. No magnetic signals were detected
in our magnetization measurements of both 3N and 6N boron
precursors before and even after heat treatment at 1,450 $\rm ^o$C
for 12 hr. We conducted our optical measurements using FTIR and
grating spectrometers covering the spectral range of 5 meV--6 eV.
For tunneling measurements we used the in-situ (breakage at 4 K)
break-junction method \cite{Ekino,Jung}, which protects the
samples from serious environmental defects.

Figure 1 displays the temperature-dependent resistivity $\rho(T)$
of CaB$_{6}$(6N) and CaB$_{6}$(3N) on a log scale. The $\rho(T)$
of CaB$_{6}$(6N) was much higher in magnitude than that of
CaB$_{6}$(3N) and exhibited typical semiconducting behavior in the
entire temperature range of 2--300 K. However, the $\rho(T)$ of
CaB$_{6}$(3N) revealed metallic temperature dependence except at
very low temperatures. The insets of Fig. 1 display the infrared
reflectivity of CaB$_{6}$(6N) and CaB$_{6}$(3N) at room
temperature. The reflectivity of CaB$_{6}$(6N) reveals its
insulating nature: the overall shape of the spectrum, dominated by
a single optical phonon mode at 150 cm$^{-1}$, is a typical
characteristic of an insulator or a pure semiconductor. We were
not able to detect a clear signature of a Drude-like feature of
free carriers down to 40 cm$^{-1}$ (5 meV) (the low-frequency
region somewhat obscured by noisy interference fringes). On the
other hand, CaB$_6$(3N) exhibited a clear Drude-like feature below
about 60 cm$^{-1}$ (7.5 meV) in addition to the aforementioned
phonon mode. Hence our optical data are consistent with the
temperature dependence of the resistivity of CaB$_6$(6N) and
CaB$_6$(3N) described above. Similar features were observed in the
transport and optical measurements on boron-deficient and
boron-rich single crystals. Our observations suggest that
boron-related defects (or chemical impurities) present in
CaB$_6$(3N) are closely associated with carrier doping and hence
with the semiconducting/metallic characteristics of the
resistivity. These results are in contrast to earlier reports in
Ref. \cite{Tsebulya} (a high-alloy semiconductor model) and
\cite{Taniguchi} (a doped semimetal model) but are compatible with
Ref. \cite{Vonlanthen} with only subtle differences. Thus, it is
essential to employ high-purity boron for careful studies of the
intrinsic properties of CaB$_{6}$. The effect of the boron purity
in alkaline-earth hexaborides has not yet been seriously addressed
despite numerous references in the literature which argue that the
experimental data on these compounds were quite sensitive to the
sample quality.

We have performed electron-tunneling experiments on CaB$_{6}$(6N)
and CaB$_{6}$(3N) using a break junction to identify the
difference in their electronic structure. The d$I$/d$V$ versus
applied voltage is plotted in Fig. 2(a) and (b) for CaB$_{6}$(6N)
and CaB$_{6}$(3N), respectively. The overall shape of the
tunneling conductance curve for CaB$_{6}$(6N) shows a well-defined
large gap structure of size 2$\Delta\approx$ 2 eV and a weak
sub-gap anomaly of size 2$\Delta^{*}\approx$ 0.4 eV. The non-zero
d$I$/d$V$ at zero bias and the broad maximum on the shoulder at
both bias indicate that the electronic states are not completely
depleted inside the large gap. The observed large gap feature
$\Delta\approx$ 1 eV is most likely a manifestation of the bulk
semiconducting band gap of pure CaB$_6$ corresponding to the 1 eV
$X$-point band gap reported in the ARPES measurements of Ref.
\cite{Denlinger} and \cite{Souma}.

We noted a drastic change in the tunneling conductance spectrum
for CaB$_{6}$(3N) as shown in Fig. 2(b). The sub-gap structure
2$\Delta^{*}$ was strongly enhanced, while the large gap feature
2$\Delta$ was sharply depressed. The marked enhancement of the
sub-gap structure is probably linked to the induced free carriers
in CaB$_{6}$(3N), but this distinct effect is not precisely in
accordance with the electron pocket picture at the \textit{X}
point observed in the ARPES measurements of Ref. \cite{Denlinger}
and \cite{Souma}. We argue that boron-related defects in
CaB$_{6}$(3N) not only induced free carriers but also created {\em
mid-gap states} at about 0.2 eV below the conduction band
\cite{cho03}. The sub-gap feature can be then naturally
interpreted as representing tunneling of the induced free carriers
at the mid-gap states across an energy gap of $\Delta^*\approx$
0.2 eV between the highest-occupied mid-gap states and the bottom
of the conduction band. This assignment also explains why the
sub-gap feature is strongly enhanced in CaB$_6$(3N) but is weakly
present in CaB$_6$(6N): the boron-related defects are far more
abundant in CaB$_6$(3N). Understandably, even the CaB$_6$(6N)
single crystals studied in this work possess such defects to some
extent, as reflected by a small trace of the sub-gap feature in
Fig. 2(a).

Clear evidence for the bulk semiconducting band gap of 1.0 eV in
CaB$_6$(6N) and for the mid-gap states at 0.18 eV below the
conduction band in CaB$_6$(3N) comes from direct optical
absorption measurements. The optical absorption coefficients of
CaB$_6$(6N) and CaB$_6$(3N) are plotted in Fig. 3. The CaB$_6$(6N)
and CaB$_6$(3N) samples exhibited optical absorption onsets at 1.0
eV and 0.82 eV, respectively. This observation directly confirms
that pure CaB$_6$ is a semiconductor with a band gap of 1.0 eV.
The band gap of 1.0 eV is consistent with our break-junction
tunneling results discussed above and coincides with the band gap
observed by ARPES \cite{Denlinger,Souma}. The red shift of 0.18 eV
in the optical absorption threshold for CaB$_6$(3N) implies that
the boron-related defects cause either band-gap narrowing or {\em
mid-gap state} formation, the latter explanation being favored by
our tunneling results above.

By combining the results of our optical and tunneling
measurements, we acquire a consistent overall picture of the
electronic structure and the band gap of CaB$_6$ as described by a
schematic diagram in the inset of Fig. 3. The energy gap $\Delta=$
1.0 eV is common to the optical absorption spectra and to the
tunneling conductance spectra of CaB$_6$(6N). The changes noted
for CaB$_6$(3N) can be understood in terms of mid-gap states
generated at 0.18 eV below the conduction band by boron-related
defects. We assign the optical absorption threshold at 0.82 eV for
CaB$_6$(3N) to the transition from the valence band to the mid-gap
states at $E_{\mbox{\tiny{VM}}}=$ 0.82 eV above the valence band.
The sub-gap feature $\Delta^*=$ 0.18 eV in
d\textit{I}/d\textit{V}, which was strongly enhanced in
CaB$_6$(3N) but weakly present in CaB$_6$(6N), represents
tunneling from the mid-gap states to the conduction band.

We also conducted isothermal magnetization measurements at 5 and
300 K for the single crystals in Fig. 1, CaB$_{6-\delta}$(6N,3N),
CaB$_{6+\delta}$(6N,3N), and also Ca$_{1-x}$La$_{x}$B$_{6}$ with
$x$ = 0.005 (6N,3N), 0.01 (6N,3N), 0.02 (3N), 0.03 (3N), and 0.04
(3N). Surprisingly, no single crystals made with 6N boron
exhibited any detectable magnetic signal in the entire temperature
range of 5--300 K. CaB$_{6}$(3N), CaB$_{6+\delta}$(3N), and
Ca$_{1-x}$La$_{x}$B$_{6}$(3N) with $x$ = 0.005, 0.01, and 0.02
revealed ferromagnetism, as can be inferred from a hysteresis loop
in the isothermal magnetization. In contrast, there was no trace
of magnetism in CaB$_{6-\delta}$(3N) and
Ca$_{1-x}$La$_{x}$B$_{6}$(3N) with $x$ = 0.03 and 0.04. As a
representative set of data, the hysteresis loops of
Ca$_{0.99}$La$_{0.01}$B$_{6}$(3N) at 5 and 300 K are plotted in
the inset of Fig. 4. We estimate the carrier density in the
relevant samples by converting the Hall resistivity data into the
effective number of free carriers using the single-carrier
(electron) model. The presence of a single carrier species in
hexaborides was already established by Hall measurements on
Eu$_{1-x}$Ca$_{x}$B$_{6}$ \cite{Rhyee,Paschen} and by ARPES
measurements on CaB$_{6}$, SrB$_{6}$, and EuB$_{6}$
\cite{Denlinger,Souma}. Figure 4 presents the most important
correlation between the saturated magnetization $M_{\mbox{\tiny
sat}}$ and the carrier density. We believe that the exotic
ferromagnetism in CaB$_6$ requires {\em simultaneous} presence of
the localized magnetic moments and free carriers within a finite
range of density (Fig. 4). Most importantly, this indicates that
the formation of the ferromagnetic state is established through
the induced free carriers occupying the mid-gap states. This fact
can possibly explain the disappearance of magnetic signal at
relatively high carrier density in Ca$_{1-x}$La$_x$B$_6$ ($x
\gtrsim 0.03$), which can cause the mid-gap states to merge or
hybridize with the conduction band.

Regarding the formation of a localized magnetic moment, it is not
clear at present whether it comes from simple magnetic impurities,
such as Fe, FeB, and Fe$_{2}$B, or is indeed associated with
boron-related defects as theoretically considered in Ref. 12.
However, it was found from micro-chemical analysis of both 3N and
6N compounds that there is no correlation between the Fe content
on the sample surface and the emergence of ferromagnetism
\cite{Park}. Because we have not detected any magnetic signals in
our magnetization measurement of both 3N and 6N boron precursors
(even after heat treatment), we certainly believe that
ferromagnetism is related to impurities, probably not magnetic, in
the boron precursor material. It will be very important to
precisely identify these impurities and the nature of the
associated defects.

In conclusion, we believe our reports are the first to
experimentally discover the creation of mid-gap states and extra
free carriers therein by boron-related defects in CaB$_6$. In
addition, we showed that the exotic ferromagnetism in CaB$_6$ has
non-magnetic-impurity origin, contrary to the previous reports. It
will be a critical issue to understand the nature of the impurity
and the relation between the mid-gap state and ferromagnetism in
CaB$_6$.\\

This work was supported by Korean Research Foundation Grant No.
KRF-2002-070-C00032 and CSCMR at SNU funded by KOSEF. The work at
Yonsei University was supported by eSSC at Postech funded by KOSEF
(J.H.K:jaehkim@phya.yonsei.ac.kr).

\newpage

\begin{figure}
\caption{Electrical resistivity $\rho$ versus temperature for
CaB$_6$, prepared with boron of 99.9999 \%\ purity (6N) (square)
and 99.9 \%\ purity (3N) (circle). Insets: Infrared reflectivity
(300 K) of CaB$_{6}$(6N) and CaB$_{6}$(3N).} \label{fig01}

\caption{Tunneling conductance d$I$/d$V$ versus applied voltage of
CaB$_{6}$, prepared with boron of (a) 99.9999 \%\ purity and (b)
99.9 \%\ purity.} \label{fig02}

\caption{Absorption coefficient versus photon energy for CaB$_{6}$
prepared with boron of 99.9999 \%\ purity and 99.9 \%\ purity. The
dotted lines are the fits to the data for a direct band gap.
Inset: schematic plot of the electronic structure and the band gap
along with  {\em mid-gap states} (VB for valence band, CB for
conduction band, and $E_{\mbox{\tiny F}}$ for Fermi level).}
\label{fig03}

\caption{Saturated magnetization $M_{\mbox{\small{sat}}}$ versus
carrier density for CaB$_6$, CaB$_{6\pm\delta}$, and
Ca$_{1-x}$La$_{x}$B$_{6}$, prepared with boron of 99.9 \%\ purity.
The solid line is to guide the eye. Inset: Hysteresis curve for
Ca$_{0.99}$La$_{0.01}$B$_{6}$ at 5 and 300 K.} \label{fig04}
\end{figure}


\begin{thebibliography}{30}

\bibitem{Young}
D. P. Young, D. Hall, M. E. Torelli, Z. Fisk, J. L. Sarrao, J. D.
Thompson, H.-R. Ott, S. B. Oseroff, R. G. Goodrich, and R. Zysler,
Nature (London) {\bf 397}, 412 (1999).

\bibitem{Zhitomirsky}
M. E. Zhitomirsky, T. M. Rice, and V. I. Anisimov, Nature (London)
{\bf 402}, 251 (1999); L. Balents and C. M. Varma, Phys. Rev.
Lett. \textbf{84}, 1264 (2000); Victor Barzykin, and Lev P.
Gor'kov, Phys. Rev. Lett. {\bf 84}, 2207 (2000).

\bibitem{Ceperley}
David Ceperley, Nature (London) \textbf{397}, 386 (1999); G.
Ortiz, M. Harris, and P. Ballone, Phys. Rev. Lett. \textbf{82},
5317 (1999).

\bibitem{Jarlborg}
T. Jarlborg, Phys. Rev. Lett. \textbf{85}, 186 (2000).

\bibitem{Massida}
S. Massida, A. Continenza, T. M. de Pascale, and R. Z. Monnier, Z.
Phys. B \textbf{102}, 83 (1997); C. O. Rodriguez, Ruben Weht, and
W. E. Pickett, Phys. Rev. Lett. \textbf{84}, 3903 (2000).

\bibitem{Tromp}
H. J. Tromp, P. van Gelderen, P. J. Kelly, G. Brocks, and P. A.
Bobbert, Phys. Rev. Lett. {\bf 87}, 016401 (2001).

\bibitem{Denlinger}
J. D. Denlinger, J. A. Clack, J. W. Allen, G.-H. Gweon, D. M.
Poirier, C. G. Olson, J. L. Sarrao, A. D. Bianchi, and Z. Fisk,
Phys. Rev. Lett. {\bf 89}, 157601 (2002).

\bibitem{Souma}
S. Souma, H. Komatsu, T. Takahashi, R. Kaji, T. Sasaki, Y. Yokoo,
and J. Akimitsu, Phys. Rev. Lett. {\bf 90}, 027202 (2003).

\bibitem{Matsubayashi}
K. Matsubayashi, M. Maki, T. Tsuzuki, T. Nishioka, N. K. Sato,
Nature {\bf 420}, 143 (2002); S. Otani and T. Mori, J. Phys. Soc.
Jpn. {\bf 71}, 1790 (2002); C. Meegoda and M. Trenary, T. Mori,
and S. Otani, Phys. Rev. B \textbf{67}, 172410 (2003).

\bibitem{Young02}
D. P. Young, Z. Fisk, J. D. Thompson, H. R. Ott, S. B. Oseroff,
and R. G. Goodrich, Nature {\bf 420}, 144 (2002).

\bibitem{Fisk}
Z. Fisk, H. R. Ott, V. Barzykin, L. P. Gor'kov, Physica B {\bf
312-313}, 808 (2002).

\bibitem{Monnier}
R. Monnier and B. Delley, Phys. Rev. Lett. {\bf 87}, 157204
(2001).

\bibitem{Ott}
H. R. Ott, M. Chernikov, E. Felder, L. Degiorgi, E. G.
Moshopoulou, J. L. Sarrao, and Z. Fisk, Z. Phys. B \textbf{102},
337 (1997).

\bibitem{Ekino}
T. Ekino, T. Takabatake, H. Tanaka, and H. Fujii, Phys. Rev. Lett.
{\bf 75}, 4262 (1995).

\bibitem{Jung}
M. H. Jung, T. Ekino, Y. S. Kwon, and T. Takabatake, Phys. Rev. B
{\bf 63}, 035101 (2001).

\bibitem{Tsebulya}
G. G. Tsebulya, G. K. Kozina, A. P. Zakharchuk, A. V. Kovalev, and
E. M. Dudnik, Powder Metallurgy and Metal Ceramics \textbf{36},
413 (1997).

\bibitem{Taniguchi}
K. Taniguchi, T. Katsufuji, F. Sakai, H. Ueda, K. Kitazawa, and H.
Takagi, Phys. Rev. B \textbf{66}, 064407 (2002).

\bibitem{Vonlanthen}
P. Vonlanthen, E. Felder, L. Degiorgi, H. R. Ott, D. P. Young, A.
D. Bianchi, and Z. Fisk, Phys. Rev. B {\bf 62}, 10076 (2000).

\bibitem{cho03}
The term ``boron-related defects" is more appropriate to stand
for the origin of the creation of the mid-gap states in
CaB$_6$(3N) because simple impurities are not likely to induce
such drastic changes in the electronic state unless they are inherently
and critically related to the chemical state of boron.

\bibitem{Rhyee}
Jong-Soo Rhyee, B. K. Cho, and H-. C. Ri, Phys. Rev. B {\bf 67},
125102 (2003).

\bibitem{Paschen}
S. Paschen, D. Pushin, M. Schlatter, P. Vonlanthen, H. R. Ott, D.
P. Young, and Z. Fisk, Phys. Rev. B {\bf 61}, 4174 (2000).

\bibitem{Park}
B. H. Park, unpublished.
\end{thebibliography}
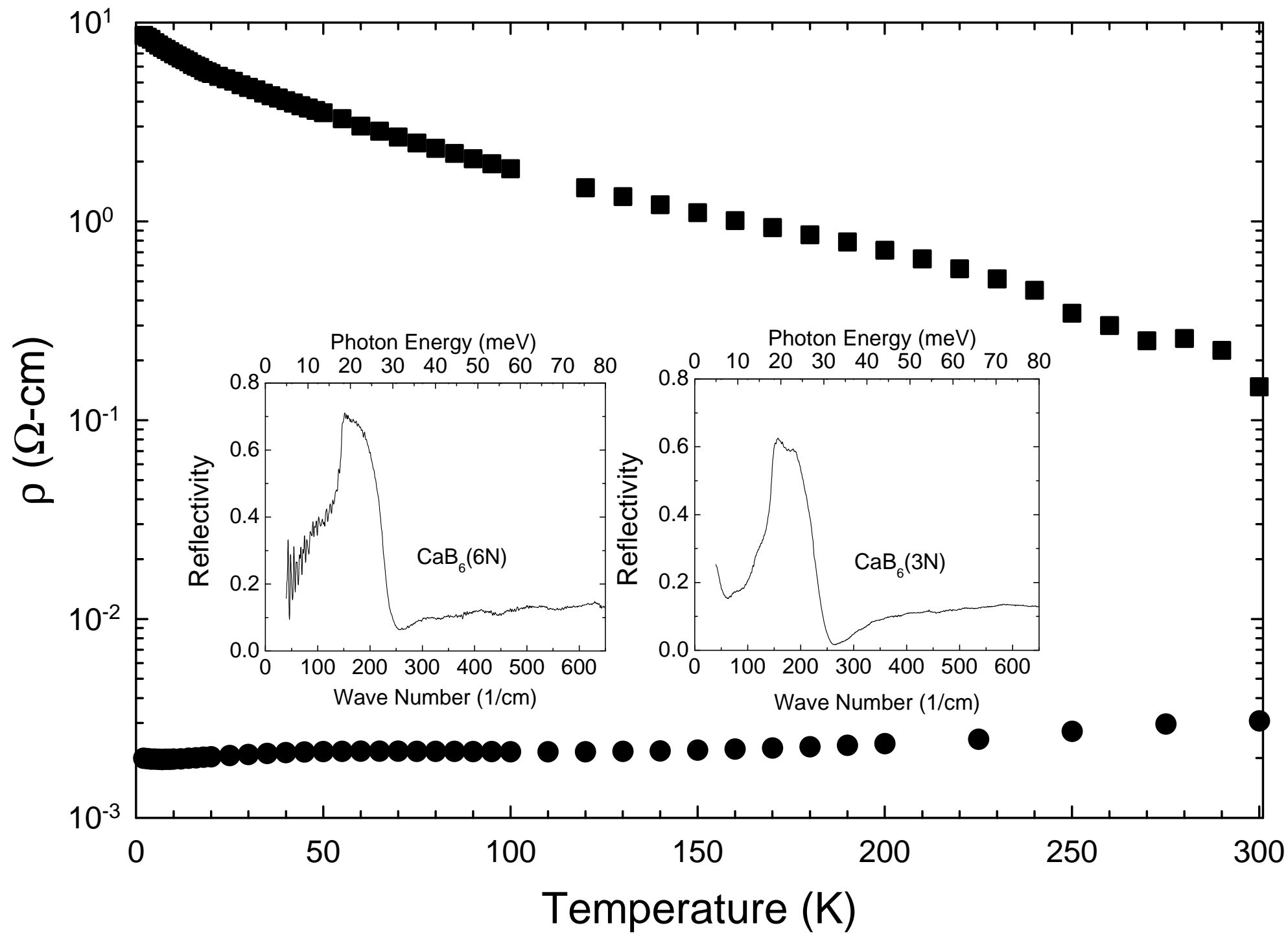
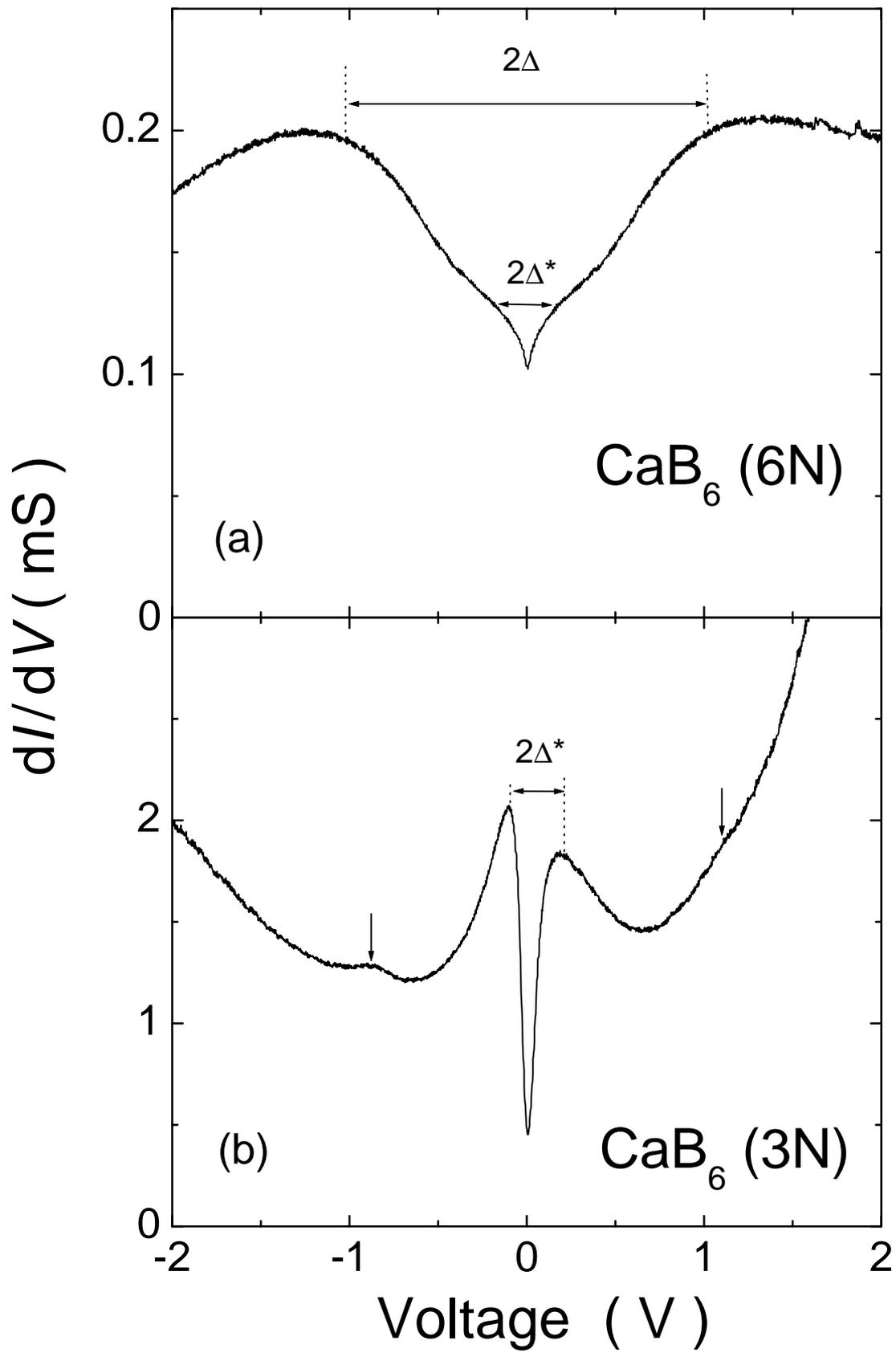
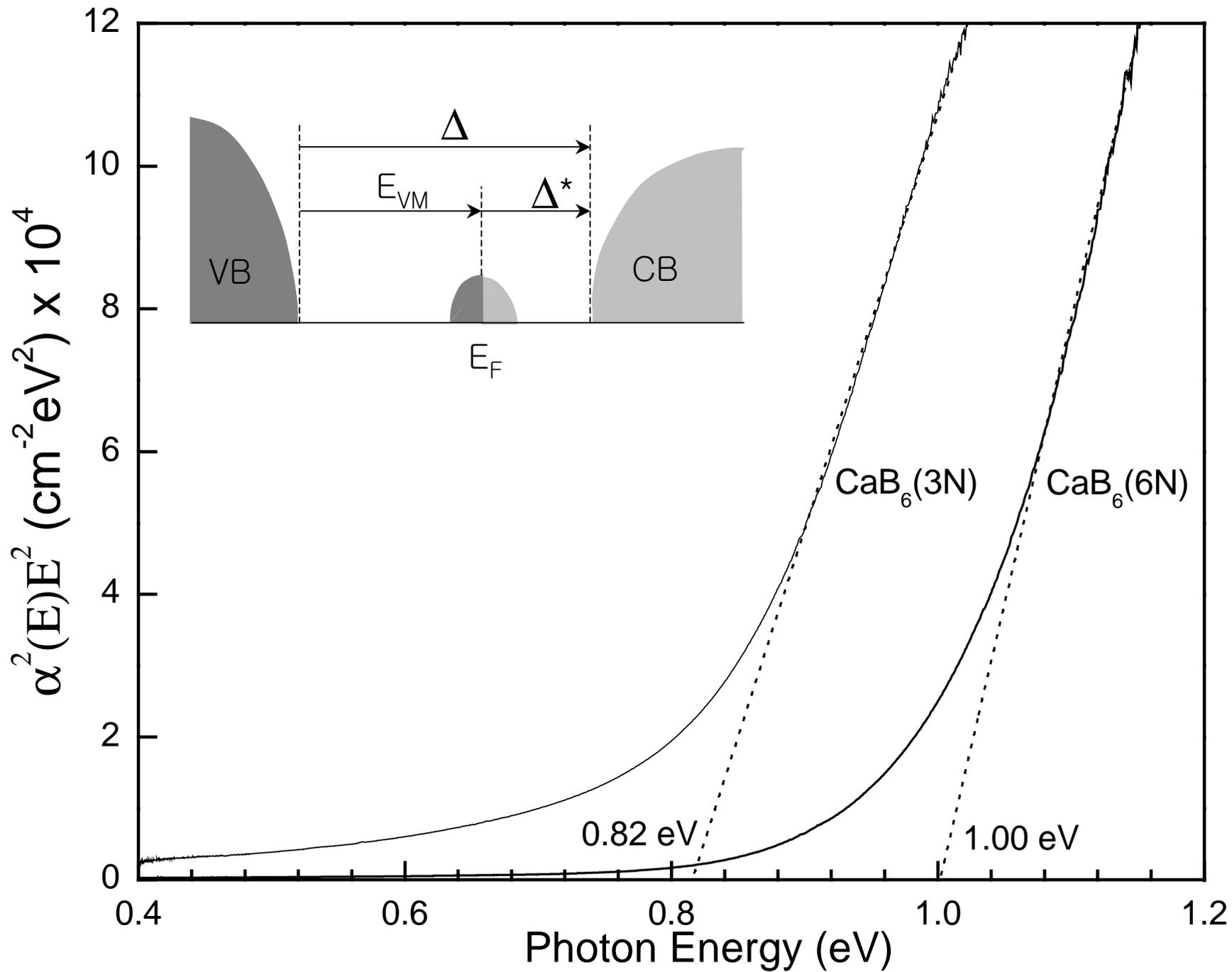
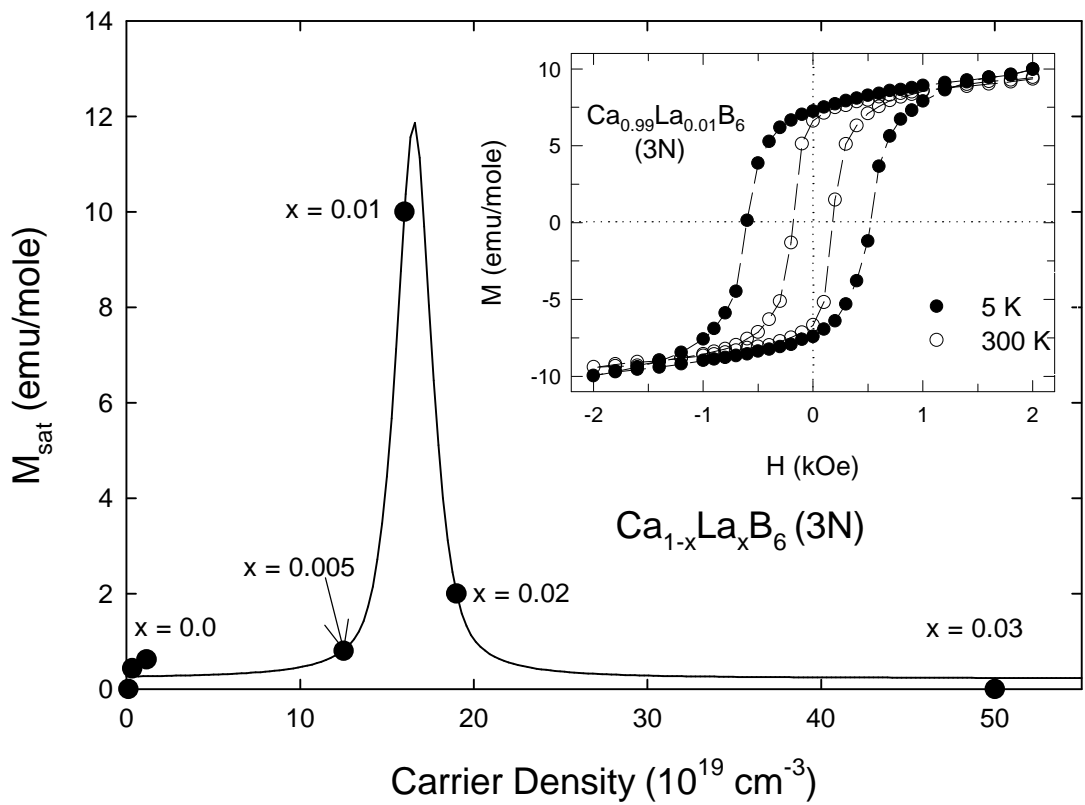
\end{document}